\documentclass[amsmath,amssymb, superscriptaddress, showpacs, floatfix,aps,prb,numerical,preprint]{revtex4-1}

\usepackage[english]{babel}
\usepackage{amssymb, amsmath}
\usepackage{bm}
\usepackage{graphicx}
\usepackage{subfigure}
\usepackage{appendix}
\usepackage{color}
\usepackage[nottoc]{tocbibind}
\usepackage{dcolumn}
\usepackage{hhline}
\usepackage{natbib}
\usepackage[siunitx]{circuitikz}
\usepackage{longtable}
\usepackage{gensymb}
\usepackage{soul}

\begin{document}

\title{Linear and planar molecules formed by coupled P donors in silicon}

\author{M.V. Klymenko}
\affiliation{Department of Chemistry, B6c, University of Liege, B4000 Liege, Belgium}

\author{S. Rogge}
\affiliation{School of Physics, The University of New South Wales, Sydney, New South Wales 2052, Australia}

\author{F. Remacle}
\email{fremacle@ulg.ac.be}
\affiliation{Department of Chemistry, B6c, University of Liege, B4000 Liege, Belgium}


%%%%%%%%%%%%%%%%%%%%%%%%%%%%%%%%%%%%%%%%%%%%%%%%%%%%%%%%%%%%%%%%%%%%%%%%%%%%%%%%%%%%%%%%%%%%

\begin{abstract}
Using the effective mass theory and the multi-valley envelope function representation, we have developed a theoretical framework for computing the single-electron electronic structure of several phosphorus donors interacting in an arbitrary geometrical configuration in silicon taking into account the valley-orbit coupling. The methodology is applied to three coupled phosphorus donors, arranged in a linear chain and in a triangle,  and to six donors arranged in a regular hexagon. The results of
the simulations evidence that  the valley composition  of the single-electron states strongly depends on the geometry of the dopant molecule and its orientation relative to the crystallographic axes of silicon.  The electron binding energy of the triatomic linear molecules is larger than that of the diatomic molecule oriented along the same crystallographic axis, but the energy gap between the ground state and the first excited state is not significantly different for internuclear distances from
1.5 to 6.6 nm. Three donor atoms arranged in a triangle geometry have larger binding energies than a triatomic linear chain of dopants with the same internuclear distances. The planar donor molecules are characterized by a strong polarization in favor of the valleys oriented perpendicular to the plane of the molecule. The polarization increases with number of atoms forming the planar molecule.  
\end{abstract}

\maketitle

\section{Introduction}

% \begin{figure}[t]
% \centering
% \subfigure[]{\includegraphics[width=7.3cm]{drawing}}
% \subfigure[]{\includegraphics[width=4.3cm]{valleys}}
% \caption{} 
% \label{fig:wf}
% \end{figure}

Single-atom transistors \cite{SAT_book} are considered as practical candidates for novel classical and quantum computing devices compatible with metal-oxide-semiconductor (MOS) technology \cite{SAT_qubit,Klein,Mol1}. Designing atomic-scale electronic nanodevices has been made possible by the recent progress in deterministic positioning of dopants in crystalline silicon \cite{SAT} and in measuring their properties by scanning tunneling microscopy (STM) \cite{SalfiVO}.  Single-atom and diatomic impurity systems have been investigated both theoretically and experimentally in terms of their single-electron energy spectra, electron density distributions and two-electron exchange coupling \cite{Abinitio, Exchange}. Going beyond a single atom by increasing the number of coupled dopants in such devices is motivated by the need to get a larger number of qubits integrated on a single die and to provide an efficient transport of electron spin qubits \cite{3QD}. In this paper, we extend those studies and computationally characterize  the electronic structure of three and six coupled donor atoms in different geometrical configurations. 

Simulating the electronic structure of several interacting dopants represents a computationally demanding multi-scale problem. The wave functions and energy spectra exhibit features both related to non-uniform electron density distribution within the unit cell and features extending up to ten nanometers. Therefore accurate modeling needs to cover lengths from atomistic to mesoscopic scales. The problem is usually tackled either by the tight-binding methodology  \cite{TB} or by means of the effective mass theory \cite{PhysRevB.84.155320, Pica}. The first one is more rigorous in treating atomistic details of wave functions, while the effective mass theory requires less computational time and has better scaling with the size of the system.

However, since its application to donors in early days, the effective mass theory has been faced with including the effect of the valley-orbit coupling  \cite{Baldereschi, Twose}. The valley-orbit coupling, induced by a fast varying part of the confinement potential, lifts the six-fold degeneracy of the conduction band minima in the presence of a fast-varying confinement potential and leads to a specific valley composition for each bound state. The valley composition of the ground state has practical implications for valleytronics \cite{valleytronics}, a new technology which exploits valley composition, that can be read as a momentum composition, as a new degree of freedom in addition to spin and charge to control electric current. Studying this phenomena for structures of many coupled donors opens new ways to enhance the valley polarization. As has been recently demonstrated, the information on the valley composition of a single and of several coupled donors is experimentally accessible  by STM  measurements \cite{SvenVO, SalfiVO}. 

 An accurate interpretation of STM images of dopants usually starts with computing their single-electron wave functions which are employed in further analysis based on the quasi-particle picture \cite{Quasiparticle}. Also, these functions may serve as a basis set for the configuration interaction method allowing to compute the many-electron wave function \cite{FCI, FCI1} or for the Hubbard model that takes into account strong correlation phenomena in coupled donor atoms \cite{Hubbard1}.

Our aim is to compute the single-electron electronic structure of several coupled phosphorus donors. We focus on the effects of valley-orbit coupling and valley composition in those systems and on ways to engineer such structures for classical and quantum computations. To do so, in Sec. II we extend our formalism, based on the multi-valley envelope function representation for a single donor atom \cite{Kly}, to a system of several coupled impurities. In Sec. III A, we describe general
properties of the valley-orbit coupling in linear and planar dopant molecules. As illustrative examples, we consider the electronic structure of three coupled phosphorus donors, arranged in a linear chain and in a triangular structure  (Sec. III B), and of six donors arranged in a hexagonal structure  (Sec. III C). All these structures are representatives of either linear or planar molecules. For the triatomic structures, we focus on the dependence of energy spectra on geometrical parameters and compare results with the single-electron spectra of two coupled phosphorus donors \cite{us}. For the hexagonal structure we study in details the effect of molecular orientation relative to the crystallographic axes of silicon and the effect of weak disorder on the electronic structure. The results are summarized in Sec. IV.

\section{Model}

The problem to be solved is represented by the stationary Schr\"{o}dinger equation with the Hamiltonian of the form $H=T+V_{Si}(\mathbf{r})+\sum_{\alpha} V(\mathbf{r}-\mathbf{r}_{\alpha})$, where $T$ is the kinetic energy, $V_{Si}(\mathbf{r})$ is the periodic potential of the silicon lattice, $V(\mathbf{r}-\mathbf{r}_{\alpha})$ is the screened potential of an impurity atom placed at the position $\mathbf{r}_{\alpha}$ and the index $\alpha$ runs over all dopants.  The dielectric function
responsible for the static screening is taken from Ref. \cite{Pantelides} where its analytical model has been derived by Pantelides and Sah basing on numerical computations of Nara \cite{Nara}. 

Since the band structure of crystalline silicon has six equivalent conduction band minima, this problem can be efficiently solved using $k \cdot p$-theory with the envelope function approximation in close vicinity of a special point in the Brillouin zone. In Ref. \cite{Kly} this approach has been extended for the case of several special points giving rise to the multi-valley envelope function representation. According to that method, the first Brillouin zone of the  crystalline silicon has to be divided into six regions, each containing a single conduction band valley. After partitioning the Brillouin zone into six sectors, the single-electron wave functions of an impurity atom can be represented as follows:

\begin{equation}
    \psi_j(\mathbf{r}) =  \sum\limits_{n,\mathbf{k}_0} f_{n,j}(\mathbf{k}_0,\mathbf{r}) u_{n,\mathbf{k}_0}(\mathbf{r})e^{i \mathbf{k}_0 \mathbf{r}},
    \label{wf}
\end{equation}
where $f_{n,j}(\mathbf{k}_0,\mathbf{r})$ is a slow varying envelope function of the $j$-th state of the impurity atom, defined for an energy band with the band index $n$ and a region of the Brillouin zone specified by the wave vector $\mathbf{k}_0$ (which is also called the valley index in the subscript notation), $u_{n,\mathbf{k}_0}(\mathbf{r})$ is a  periodic Bloch function. Each of six valleys specified by the vector $\mathbf{k}_0$ can be also designated as $\{-X,X,-Y,Y,-Z,Z\}$ where the
letters correspond to the axes along which the isoenergetical ellipsoids of the valleys are oriented.  
 
The expression (\ref{wf}), called the envelope function representation, is quite general and allows for both band mixing and valley mixing \cite{Foreman_mix}. From now on we use the single band approximation and drop the index $n$. In this representation the Schr\"{o}dinger equation reduces to  a system of six coupled eigenvalue problems. At this stage the $k \cdot p$-method with the effective mass approximation has been applied to each of them separately leading to the system of envelope function equations \cite{Kly}:

\begin{eqnarray}
    \left[H_{kp}(\mathbf{k}_0, \mathbf{k} \rightarrow i\nabla)
        +\sum\limits_{\alpha} V_{\mathbf{k}_0,\mathbf{k}_0}(\mathbf{r}-\mathbf{r}_{\alpha})\right]f(\mathbf{k}_0,\mathbf{r})\nonumber \\
        +\sum\limits_{\alpha,\mathbf{k}_0'\neq \mathbf{k}_0} V_{\mathbf{k}_0',\mathbf{k}_0}(\mathbf{r}-\mathbf{r}_{\alpha}) f(\mathbf{k}_0',\mathbf{r})=Ef(\mathbf{k}_0,\mathbf{r}).
\label{final_sys}
\end{eqnarray}

%% e^{i\left(\mathbf{k}_0'- \mathbf{k}_0 \right)\mathbf{r}_{\alpha}}

The $k \cdot p$ procedure affects only the kinetic energy term and does not affect the potential energy term. In Eq. (\ref{final_sys}), the potential energy is expanded into two terms: one is diagonal in terms of valley indices and the other is non-diagonal. The equations for different valleys are coupled by the non-diagonal potential energy term \cite{Kly}:

\begin{widetext}
\begin{equation}
    V_{\mathbf{k}_0,\mathbf{k}'_0}(\mathbf{r}-\mathbf{r}_{\alpha})=\int d\mathbf{r}'' u_{n}^{*}(\mathbf{k}_0,\mathbf{r}'')V(\mathbf{r}'')u_{n}(\mathbf{k}_0',\mathbf{r}'') \Delta_{\mathbf{k}_0}(\mathbf{r} - \mathbf{r}'')e^{i(\mathbf{k}'_0-\mathbf{k}_0)\mathbf{r}''}.
    \label{pot01}
\end{equation}
\end{widetext}
where $\Delta_{\mathbf{k}_0}(\mathbf{r} - \mathbf{r}'')$ is a low-pass filter function that results from the inherent constraints imposed on the envelope function (see Ref. \cite{Kly} for the detailed derivation). 

For the case $\mathbf{k}_0 \neq \mathbf{k}'_0$, the effective potentials are very localized in the so-called central cell region. The central cell is small enough that variations of the envelope functions within that region can be neglected. Assuming that central cells of neighboring dopants do not overlap, the integral may be rewritten in the coordinate system where the origin is at the nucleus of the dopant:  

\begin{widetext}
\begin{equation}
    V_{\mathbf{k}_0,\mathbf{k}'_0}(\mathbf{r}-\mathbf{r}_{\alpha})=e^{-i\left(\mathbf{k}_0'- \mathbf{k}_0 \right)\mathbf{r}_{\alpha}}\int d\mathbf{r}'' u_{n}^{*}(\mathbf{k}_0,\mathbf{r}'')V(\mathbf{r}'')u_{n}(\mathbf{k}_0',\mathbf{r}'') \Delta_{\mathbf{k}_0}(\mathbf{r} - \mathbf{r}'')e^{i(\mathbf{k}'_0-\mathbf{k}_0)\mathbf{r}''}.
    \label{pot1}
\end{equation}
\end{widetext}
The periodic Bloch functions of bulk silicon in Eq. (\ref{pot1}) have been computed using DFT-LDA PAW method implemented in ABINIT software \cite{Gonze}.

Unlike in the case of a single impurity (see discussion in Ref. \cite{Kly}), in a polyatomic system the phase factor $e^{-i\left(\mathbf{k}_0'- \mathbf{k}_0 \right)\mathbf{r}_{\alpha}}$ in Eq. (\ref{pot1}) plays crucial role determining the oscillatory dependence of the electronic structure on the internuclear distances.

The integral (\ref{pot1}) has been numerically computed using the convolution theorem and fast-Fourier transform. The resulting potentials are called the effective potentials \cite{Kly}. They have several properties which allow simplifying the problem. First, at $\mathbf{k}_0=\mathbf{k}'_0$ they are similar to the Coulomb potential at distances far from the impurity nucleus, but, unlike the Coulomb potential, they do not have a singularity point and they are smooth at the nucleus.  Thus, the
effective potentials can be computed once and stored for other computations, similarly to the pseudopotentials in electronic structure computations for crystalline solids.

Eq. (\ref{final_sys}) with the effective potentials (\ref{pot1}) has been solved in two steps. First, we neglect coupling terms that lead to a system of six independent single-valley envelope-function equations. Details on the numerical method, applied for the single-valley envelope function equation, are provided in Appendix A. The obtained solutions, $\varphi_{j,\mathbf{k}_0}(\mathbf{r})$, form a full basis set for each valley. We use these solutions as a basis set for the linear variational method representing the unknown envelope functions $f(\mathbf{k}_0,\mathbf{r})$ as follows:

\begin{equation}
    f(\mathbf{k}_0,\mathbf{r}) = \sum\limits_{j} a_{j,\mathbf{k}_0} \varphi_{j,\mathbf{k}_0}(\mathbf{r}) ,
    \label{series}
\end{equation}

At the second step, substituting the expression (\ref{series}) into Eq. (\ref{final_sys}), we get a linear eigenvalue problem:

\begin{equation}
\mathbf{B}\mathbf{C}=E\mathbf{C},
    \label{lin_sys}
\end{equation}
where $\mathbf{C}$ is the vector of unknown expansion coefficients. 

Strictly speaking, the sum in the expression  (\ref{series}) should be performed also over the valley index $\mathbf{k}_0'$ which implies using a non-orthogonal basis set for each envelope function equation.  In this case, Eq. (\ref{lin_sys}) should be rewritten as $\mathbf{B}\mathbf{C}=E\mathbf{S}\mathbf{C}$, where $\mathbf{S}$ is the overlap matrix. Although this is a more general formalism, using such non-orthogonal basis set does not improve the accuracy significantly because the coupling potential is very localized. The elements of the matrix $\mathbf{B}$  in Eq. (\ref{lin_sys}) reads: 
\begin{equation}
    B_{\mathbf{k}_0,\mathbf{k}_0'}^{i,j}=\begin{cases}
    E_{j,\mathbf{k}_0}^s \delta_{i,j}, & \text{if } \mathbf{k}_0=\mathbf{k}_0';\\
    M_{\mathbf{k}_0,\mathbf{k}_0'}^{i,j}, &\text{if } \mathbf{k}_0 \neq \mathbf{k}_0',
    \end{cases}
\end{equation}
where $E_{j,\mathbf{k}_0}^s$ is the eigenvalue of the single-valley envelope-function equation,

\begin{eqnarray}
M_{\mathbf{k}_0,\mathbf{k}_0'}^{i,j}&&=\sqrt{6}\sum_{\alpha}e^{-i\left(\mathbf{k}_0'- \mathbf{k}_0 \right)\mathbf{r}_{\alpha}} \times \nonumber \\ 
&&\int d\mathbf{r} \varphi_{i,\mathbf{k}_0}(\mathbf{r})  V_{\mathbf{k}_0',\mathbf{k}_0}(\mathbf{r}-\mathbf{r}_{\alpha}) \varphi_{j,\mathbf{k}'_0}(\mathbf{r})
    \label{nondiag_me_s}
\end{eqnarray}
Since the effective potential for $\mathbf{k}_0' \neq \mathbf{k}_0$ is localized within the central cell and variations of the envelope functions within the central cell can be neglected, Eq. (\ref{nondiag_me_s}) reduces to:
\begin{eqnarray}
M_{\mathbf{k}_0,\mathbf{k}_0'}^{i,j}&&= \sqrt{6}\int d\mathbf{r} V_{\mathbf{k}_0',\mathbf{k}_0}(\mathbf{r}) \times\nonumber \\ 
&& \sum_{\alpha}e^{-i\left(\mathbf{k}_0'- \mathbf{k}_0 \right)\mathbf{r}_{\alpha}} \varphi_{i,\mathbf{k}_0}(\mathbf{r}_{\alpha})\varphi_{j,\mathbf{k}'_0}(\mathbf{r}_{\alpha}) 
    \label{nondiag_me_s1}
\end{eqnarray}
This approximation is known as a contact-potential approximation, and it has been  first proposed and successfully justified by Friesen \textit{et al.} for the silicon quantum wells \cite{Friesen_well} and quantum dots \cite{Friesen_dot}. The integral in the expression (\ref{nondiag_me_s1}) may either be computed numerically from the first-principles \cite{Kly} or be fitted to the experimental data for the single P donor in bulk silicon with two fitting parameters: 
\begin{equation}
\sqrt{6}\int d\mathbf{r} V_{\mathbf{k}_0',\mathbf{k}_0}(\mathbf{r}) \approx  \alpha \delta_{\mathbf{k}_0=-\mathbf{k}'_0}+\beta \delta_{\mathbf{k}_0  \bot \mathbf{k}'_0}
\end{equation}
with $\alpha=-0.99$ meV and $\beta=-1.72$ meV.

\section{Results}

\subsection{General properties of linear and planar P molecules in silicon}

We begin by providing a general analysis based on symmetry considerations that allows to analyse the numerical results. As was shown in Refs. \cite{Kly, Voon}, the envelope function representation can be thought of as a linear combination of states taken from energy bands of bulk silicon combined with the $k \cdot p$ perturbation technique. For shallow donors, the dominating terms in such a linear combination are the quantum states of silicon taken from the vicinity of six equivalent  conduction band minima.  For a single donor atom, the valley population of each quantum state can be explained using group theory analysis applied to the tetrahedral symmetry of the central cell potential. Such a qualitative analysis is confirmed by numerical computations based on diagonalizing matrix $\mathbf{B}$ defined by Eq. (\ref{lin_sys}).

In the case of several coupled donors, the symmetry of the central cell is superimposed on the symmetry of the dopant molecule which is reflected in envelope functions and the phase factors in the Eq.  (\ref{nondiag_me_s1}). Let us analyse Eq. (\ref{nondiag_me_s1}) for the two coupled donors placed at the coordinates $-x$ and $x$ at the internuclear axis oriented along the crystallographic axis [100] with the internuclear distance $l=2x$. To make the qualitative analysis simpler, we consider the valley-orbit coupling only for the orbits with indices $j=i=1$ with an even envelope function $\varphi_{1,\mathbf{k}_0}(-x)=\varphi_{1,\mathbf{k}_0}(x)$.

It follows from Eq.  (\ref{nondiag_me_s1}) that, compared to the case of a single donor, all matrix elements $M_{\mathbf{k}_0,\mathbf{k}_0'}^{1,1}$ remain unchanged except those involving valleys $-X$ and $X$ oriented along the axes [100]. For those valleys, the coeffcients read:
\begin{eqnarray}
M_{\mathbf{k}_0= \{ \pm X\},\mathbf{k}_0' \neq \{ \pm X\}}^{1,1}=2 \cos \left(|\mathbf{k}_0| x \right) \varphi_{1,X}(x)\varphi_{1,Y}(x) \int d\mathbf{r} V_{\mathbf{k}_0',\mathbf{k}_0}(\mathbf{r}) 
\label{two1}
\end{eqnarray}
and
\begin{eqnarray}
M_{\mathbf{k}_0= \{ \pm X\},\mathbf{k}_0'|| \mathbf{k}_0}^{1,1}=2 \cos \left(|\mathbf{k}_0| l \right) |\varphi_{1,X}(x)|^2 \int d\mathbf{r} V_{\mathbf{k}_0',\mathbf{k}_0}(\mathbf{r}).
\label{two2}
\end{eqnarray}

The splitting of energy states, containing the valleys $-X$ and $X$, is characterized by an oscillatory dependence on the internuclear distance, while states whose valley composition exclude these valleys have a smooth dependence on the internuclear distance. This makes the states composed from the valleys $-X$ and $X$ sensitive to small variations in positioning of donor atoms relative to each other. Also, by choosing the proper distance between donor atoms, one may reduce the contribution of
the valleys  $-X$ and $X$ in to the ground state. This effect can be considered as a resonance phenomenon when electron waves interfere at the dopants in the molecules.  The effect can be exploited to engineer the valley composition of  donors quantum states. Let us place a third P atom at the coordinate $0$ to form a chain of three dopants oriented along the axis [100] having a symmetrical envelope function relative to the center atom. In this case Eqs. (\ref{two1}) and
(\ref{two2}) are modified as follows:

\begin{eqnarray}
M_{\mathbf{k}_0= \{ \pm X\},\mathbf{k}_0' \neq \{ \pm X\}}^{1,1}=\left[\varphi_{1,X}(0)\varphi_{1,Y}(0)+ 2 \cos \left(|\mathbf{k}_0| x \right) \varphi_{1,X}(x)\varphi_{1,Y}(x) \right] \int d\mathbf{r} V_{\mathbf{k}_0',\mathbf{k}_0}(\mathbf{r}) 
\label{three1}
\end{eqnarray}
and
\begin{eqnarray}
M_{\mathbf{k}_0= \{ \pm X\},\mathbf{k}_0'|| \mathbf{k}_0}^{1,1}=\left[ |\varphi_{1,X}(0)|^2+ 2 \cos \left(2|\mathbf{k}_0| x \right) |\varphi_{1,X}(x)|^2 \right] \int d\mathbf{r} V_{\mathbf{k}_0',\mathbf{k}_0}(\mathbf{r}).
\label{three2}
\end{eqnarray}

In Eqs. (\ref{three1}) and (\ref{three2}), there is a large contribution in the valley-orbit coupling from the constant term proportional to the electron density at the nucleus of the central atom  $|\varphi_{1,X}(0)|^2$. The amplitude of the oscillations is proportional to the electron densities at the most left and most right atoms, $|\varphi_{1,X}(\pm x)|^2$, of the triatomic chain. The value of $|\varphi_{1,X}(\pm x)|^2$ is smaller in the three-atomic chain than in the two coupled atoms due to the
normalization of the overall envelope function which leads to a redistribution of electron density over all three atoms. Thus, the dopant molecule may be engineered in a way to reduce its sensitivity with respect to the spatial disorder.

For a single donor atom, all six valleys contribute equally into the ground state. Since the cosine factors in Eqs. (\ref{two1}) -(\ref{three2}) take values from zero to one, the contribution of $\pm X$-valleys into the ground state of two coupled donors can only be reduced or remain unchanged compared to the case of a single dopant. The contributions from several valleys can be decreased simultaneously in planar structures. For instance, a rectangular dopant molecule formed by P donor atoms with coordinates $(-x,-y)$, $(-x,y)$, $(x,-y)$ and $(x,y)$ has following valley-orbit coupling coefficients:

\begin{eqnarray}
M_{\mathbf{k}_0= \{\pm X,\pm Y\},\mathbf{k}_0' \neq \{\pm X,\pm Y\}}^{1,1}=4\frac{\cos \left(|\mathbf{k}_0| x \right)+\cos \left(|\mathbf{k}_0| y \right) }{2} \varphi_{1,X}(x)\varphi_{1,Y}(x) \int d\mathbf{r} V_{\mathbf{k}_0',\mathbf{k}_0}(\mathbf{r}) 
\label{four1}
\end{eqnarray}
and
\begin{eqnarray}
M_{\mathbf{k}_0= \{\pm X,\pm Y\},\mathbf{k}_0'|| \mathbf{k}_0}^{1,1}= 4\frac{\cos \left(2|\mathbf{k}_0| x \right)+\cos \left(2|\mathbf{k}_0| y \right) }{2} |\varphi_{1,X}(x)|^2 \int d\mathbf{r} V_{\mathbf{k}_0',\mathbf{k}_0}(\mathbf{r}).
\label{four2}
\end{eqnarray}

In Eqs. (\ref{four1}) and (\ref{four2}), we have assumed that the ground state envelope function has equal values at all four donor atom nuclei placed in the plane perpendicular to the crystallographic axis [001]. The fractions in  Eqs. (\ref{four1}) and (\ref{four2}) can take values in the interval from zero to one. Thus, by choosing a proper positioning of donor atoms, we may totally filter out the valleys $\pm X$ and $\pm Y$. In planar molecules with larger number of atoms, it is nearly
impossible to find such a configuration of atoms that all phase factors equal one. Therefore, we may generally conclude that the planar donor molecules are characterized by a strong polarization in favor of the valleys oriented perpendicular to the plane of the molecule.

\subsection{Three coupled donors}

We begin by investigating the role of the VO coupling on the electronic spectrum of triatomic dopant molecules (both in linear and in triangular geometry) and compare it to the diatomic case. We focus first on the energy spectra as a function of the distance between the dopant atoms.

\begin{figure*}[t]
\centering
\includegraphics[width=17cm]{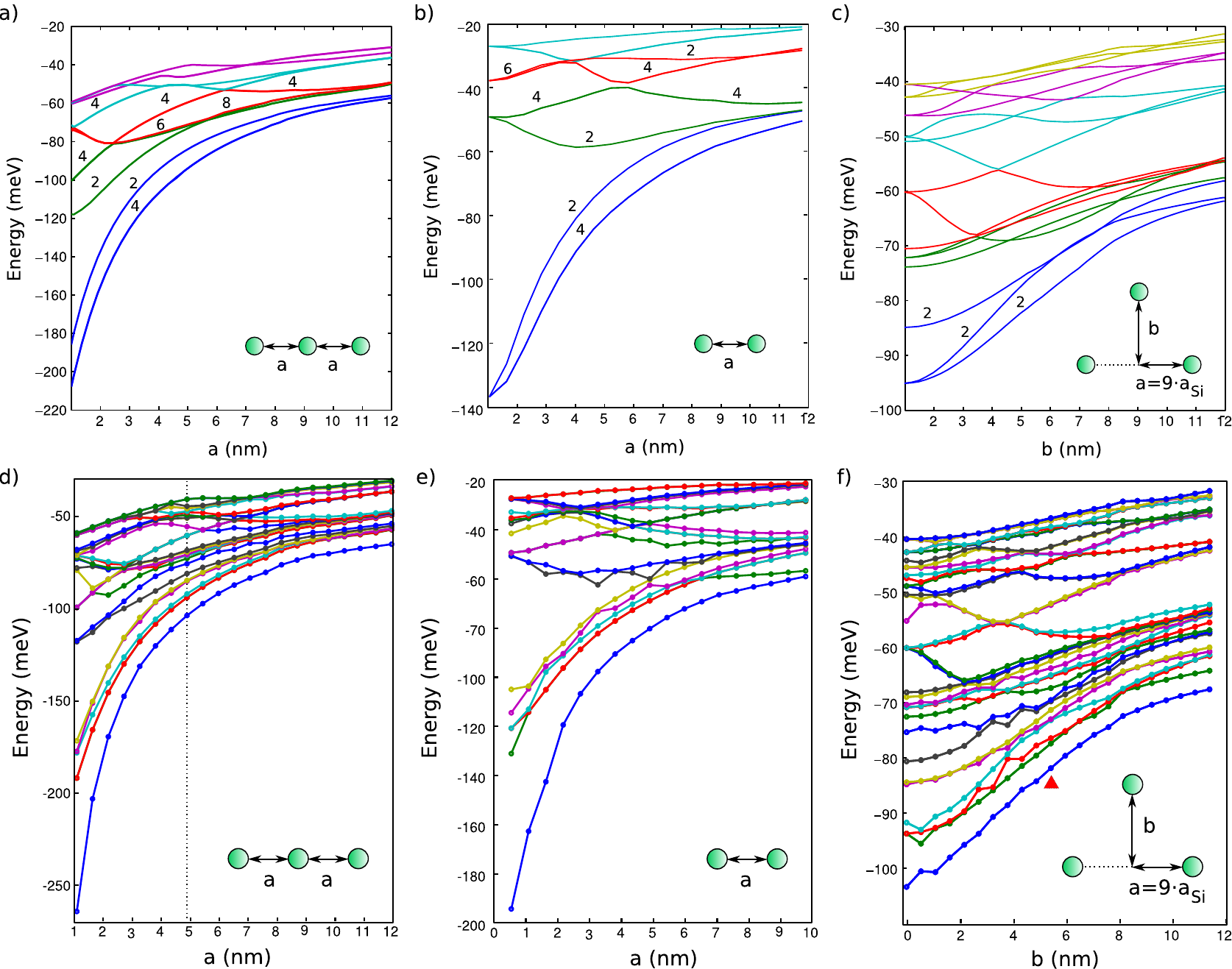}
\caption{Electronic energy spectra for  a), d) three phosphorus donors arranged in a linear chain b), e), two phosphorus donors and c), f) three phosphorus donors arranged in a triangular structure. Panels a), b), c) represent solutions of single-valley problems neglecting the valley-orbit coupling, while panels d), e), f) are for results with the valley-orbit coupling taken into account. Both diatomic and triatomic linear molecules are oriented along the crystallographic axis [100]. Numbers
above curves indicate the degrees of degeneracy of corresponding energy levels. The dotted line in d) corresponds to the energy spectrum in f) at $b=0$. The red filled triangle in the panel f) designates the ground state energy that corresponds to the wave function shown in Fig. \ref{fig:wf3_2}}.
\label{el_spec}
\end{figure*}

All panels in the upper row of Fig. \ref{el_spec}  correspond to computations where the valley-orbit coupling is neglected, while it is taken into account in the results presented in the lower row. Electronic energy spectra for three equally spaced donor atoms aligned along crystallographic axis [100] are shown in Fig. \ref{el_spec}a and \ref{el_spec}d as a function of the internuclear distance. For comparison, we also show in Fig. \ref{el_spec}b and \ref{el_spec}e the computed electronic
spectra for a diatomic donor molecular ion aligned along the same axis. Detailed information specific to the diatomic donor molecular ion can be found in Ref. \cite{us}.  To analyze the electronic spectra of three donors arranged in several triangle configurations we start with a linear structure having an internuclear distance of nine lattice constants (this point is marked as a dotted line in Fig. \ref{el_spec}d) and move the central atom perpendicular to the internuclear axis in the plane
(001) as is illustrated in Figs. \ref{el_spec}c and \ref{el_spec}f. The dot markers on all curves correspond to physical positions of the donors at sites of the silicon crystal lattice. The small distortion of the lattice caused by the impurities has been neglected. Placing the impurities between sites of the crystal lattice has no physical meaning, the lines connecting the dots are a guide to the eye along a given energy level.

In Figs. \ref{el_spec}a,d and in Figs. \ref{el_spec}b,e the energy spectra asymptotically tend to the spectrum of three and two independent donors respectively. In both cases at a distance of about 12 nm, the energy levels have not reached  yet asymptotic values and their energies continue increasing, although the electronic coupling is small. This weak coupling is due to long-range electrostatic coupling between impurities. For the triatomic structures, the long-range electrostatic interaction
is more pronounced. It causes the splitting between the ground state and the first excited state at large internuclear distances; in the single-electron spectra, in the limit of uncoupled donors the ground state is degenerate with a degeneracy equal to the number of impurities.

In the triangle structures (Figs. \ref{el_spec}c,f), the spectra exhibit  two asymptotes: one is for the single phosphorus donor (upper manifold of states) while the lower states correspond to the energy spectrum of the diatomic donor molecular ion with the internuclear distance of 18 lattice constants. The upper asymptotic manifold of states corresponds to the wave function localized on a single donor. The splitting of this asymptote into three bands instead of three exactly degenerate states as is the case for a single dopant is due to the electrostatic coupling of the dopant atom with the two others. 

Comparing the energy spectra of Figs. \ref{el_spec} a,b,c and Figs. \ref{el_spec} d,e,f, we note that even when the valley-orbit coupling is neglected, one observes a lifting of the six-fold degeneracy that is due to the effective mass anisotropy (see Fig. \ref{el_spec}, the upper row). Since the diatomic and triatomic linear structures have the same orientation relative the crystallographic axes, their spectra are quite similar. The valley-orbit coupling partially removes the
degeneracy of the energy levels and leads to an oscillatory dependence of some energy levels as a function of the internuclear distance (see the discussion in the previous section). Which state has an oscillatory dependence is determined by the orbital symmetry of its envelope functions, valley composition and by the molecule alignment relative to the crystallographic axes. The ground state is non-degenerate for all cases due to the valley-orbit coupling at small internuclear distances and due to the electronic coupling between impurities at large internuclear distances.

\begin{figure}[t]
\centering
\includegraphics[width=8.6cm]{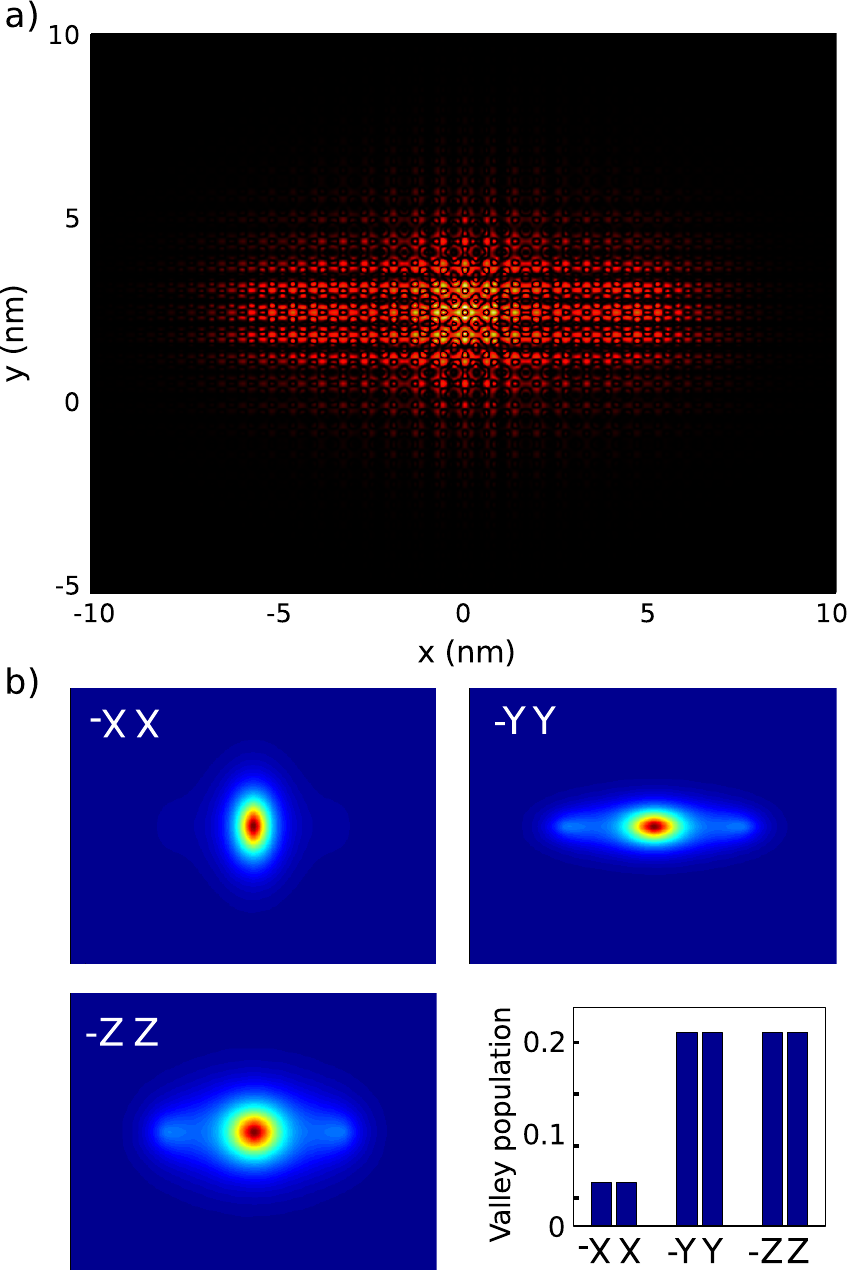}
\caption{a) The ground state electron density for the linear triatomic structure and b) corresponding single-valley envelope functions and valley compositions. The energy spectra of the linear chain is marked by the dashed line Fig. \ref{el_spec}d.} 
\label{fig:wf3_1}
\end{figure}

\begin{figure}[t]
\centering
\includegraphics[width=8.6cm]{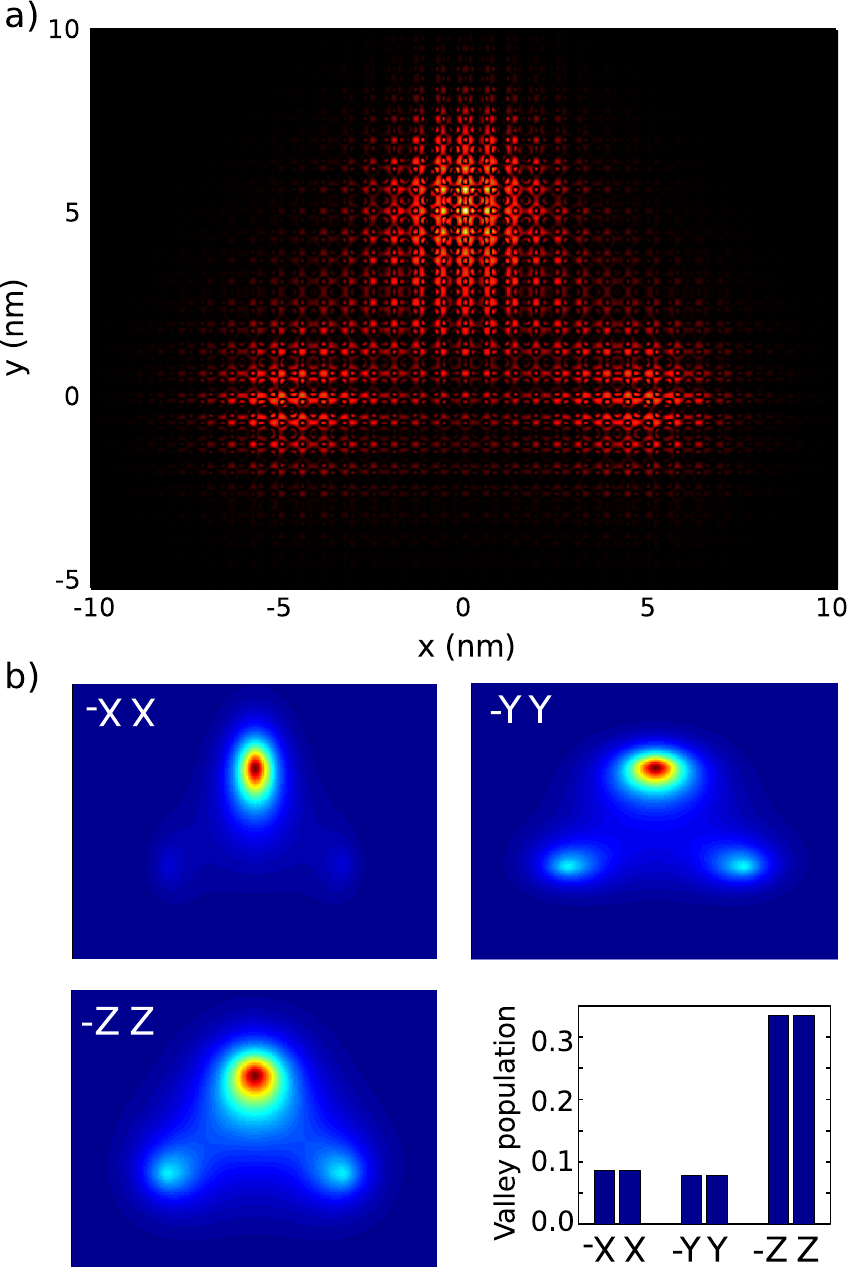}
\caption{a) The ground state electron density for the triangular structure and b) corresponding single-valley envelope functions and valley compositions. The energy spectra of the triangular molecule is marked by the red triangle in Fig. \ref{el_spec}f.} 
\label{fig:wf3_2}
\end{figure}

The computed ground state electron densities and corresponding valley populations of donors arranged in the linear and triangular structures are shown in Figs. \ref{fig:wf3_1} and \ref{fig:wf3_2}. According to Eq. (\ref{nondiag_me_s1}), the valley composition of each state depends on the overlap of the corresponding single-valley envelope function with the central-cell potential. Analyzing the single-valley envelope functions shown in the lower panels in Figs. \ref{fig:wf3_1} and
\ref{fig:wf3_2}, we conclude that due to the effective mass anisotropy some envelope functions, associated with specific valleys, are characterized by a better overlap with  the Coulomb potential of the donors and their central-cell potentials that enhances their contribution from the corresponding valleys. As a result, the ground state wave functions are characterized by larger contribution from the valleys $\pm$Y and $\pm$Z for the linear molecule (see Fig. \ref{fig:wf3_1}), while in the case of the triangular structure the ground state is polarized in favor of the $\pm$Z valleys (see Fig. \ref{fig:wf3_2}).  In the linear molecule, the valleys $\pm$X have the smallest contribution because the electron effective mass along the molecular axis for that valley is large and the wave function is strongly localized with its maximum at the central atom, while for all other valleys the effective mass in that direction is small that leads to better overlap of the single-valley envelope function with all three central-cells. In the triangle structure, there are only $\pm$Z valleys, oriented perpendicular to the plane of the triangle, whose envelope functions have relatively large magnitudes at all three impurity nuclei simultaneously. 

\begin{figure}[t]
\centering
\includegraphics[width=8.0cm]{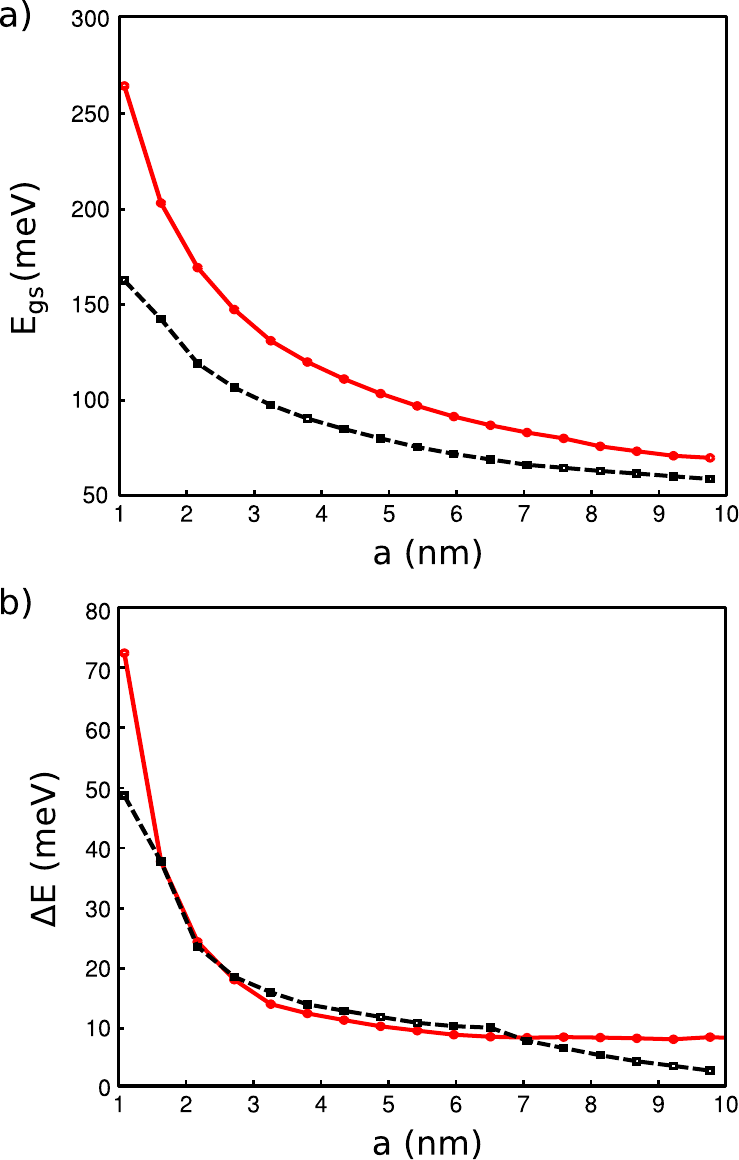}
\caption{a) Ionization energies and b) energy gaps between the ground state and the first excited state for the triatomic linear structure of donors (red line with round markers) and the diatomic structure (black line with rectangular markers).} 
\label{fig:en}
\end{figure}

In Fig. \ref{fig:en} we report on ionization energies and energy gaps between the ground state and the first excited state, data that are relevant for engineering physical implementations of qubits or classical computations on phosphorus donors. The computed ionization energy is larger for the triatomic linear structure than for the diatomic donor ion since the latter corresponds to a shallower potential well for the electron. However, the energy gap between the ground state and the first excited state is about the same for the two structures at internuclear distances from 1.5 to 6.6 nm. For the distances up to  6.6 nm, the splitting between energy levels is dominated by the valley-orbit coupling implying that it depends on the overlap of the envelope functions and the central cell potentials for all impurities according to Eq. (\ref{nondiag_me_s1}). Adding a new impurity in the system brings a new central cell potential, however the magnitude of the envelope function is redistributed  over all impurities due to its normalization. As a result, the total overlap integral is almost unchanged. The situation is different at distances below 1.5 nm, where the value of envelope function at the central cell is dramatically affected by an additional impurity placed nearby in a nonlinear way. For distances larger than 6.6 nm, the energy gap between the ground state and the first excited state is also very sensitive to  adding an additional impurity and is determined by the weak electronic coupling \cite{Exchange} and long-range electrostatic coupling of the positively charged phosphorus nuclei. 

The binding energy of three donors arranged in a triangle structure is larger than the binding energy of the triatomic linear chain at equal internuclear distances. For instance, the ground state energy for the molecule whose shape is close to an equilateral triangle with an edge length of 18 lattice constants (the base of the triangle is equal 9.77 nm and $b=7.1$ nm) equals -75.91 meV, while the ground state energy of the triatomic linear molecule having the internuclear distances of 9.77 nm equals -70.01 meV.

\subsection{Six coupled donors in hexagonal structures and effect of disorder}

We now turn to the case of six interacting dopants in a hexagonal geometry. Since there are several ways to arrange them with respect to the Silicon crystal lattice, we use a nomenclature that uniquely specifies the arrangement of any number of impurities in a regular polygon structure in the silicon crystal lattice. The nomenclature, for instance, reads P$_613_{\{100\}}$: here the letter P stands for phosphorus, the first index specifies number of impurities forming a regular polygon, the next number indicates the total number of atoms along one of edges, and the last index indicates a crystallographic axis along which this edge is aligned. More information on possible ways to arrange six atoms into a regular hexagonal structure is given in Appendix B. Here we focus on two particular hexagonal structures, P$_613_{\{100\}}$ and P$_617_{\{110\}}$, which have almost equivalent edge lengths (6.52 nm and 6.14 nm respectively), but different orientations relative to the crystallographic axes of silicon.

\begin{figure}[!t]
\centering
\includegraphics[width=8.6cm]{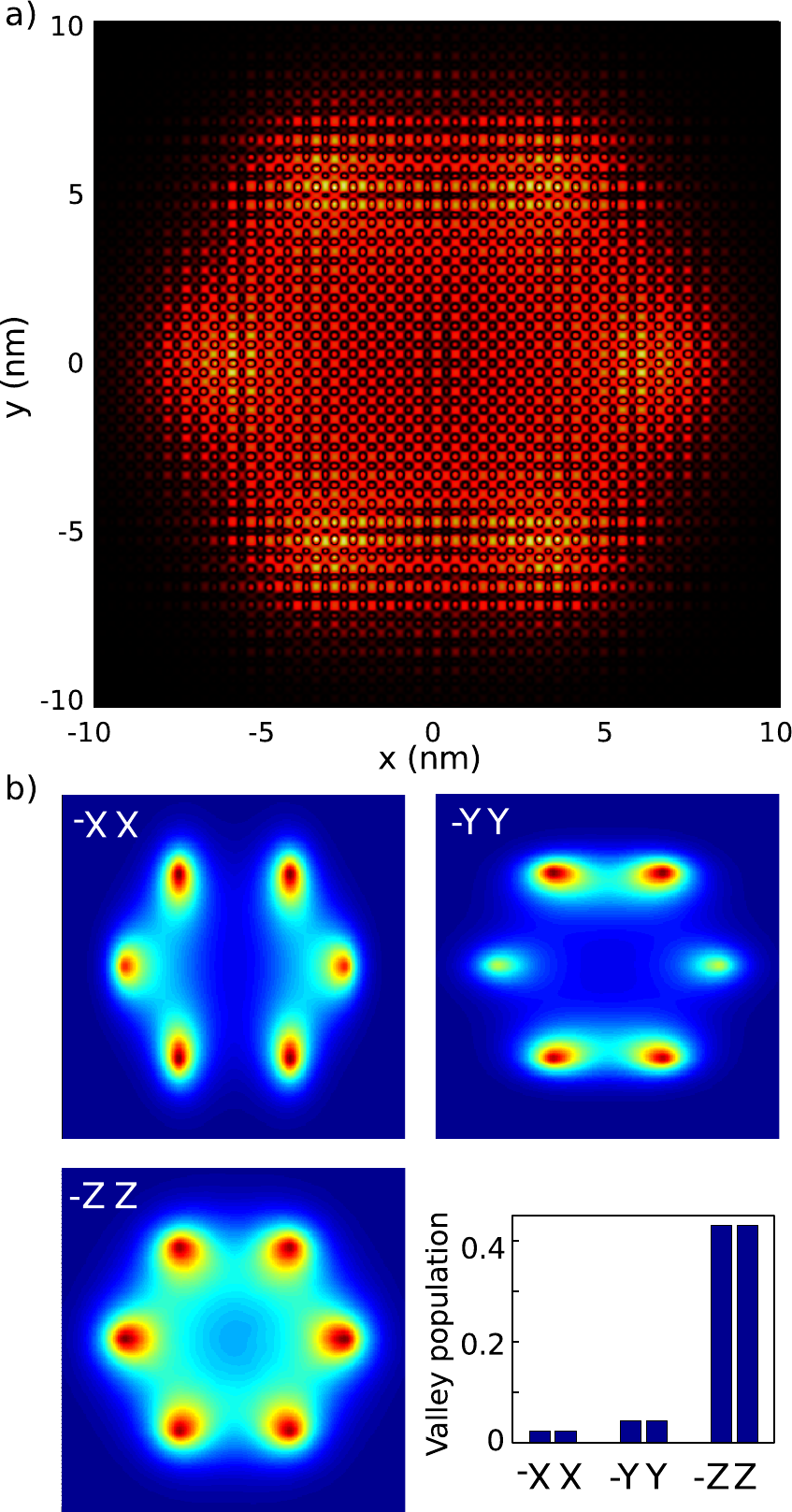}
\caption{a) The ground state electron density for the hexagonal structure formed by six donor atoms with an edge of the hexagon aligned long [100] crystallographic axis and b) corresponding single-valley envelope functions and valley compositions.} 
\label{fig:hex_1}
\end{figure}

\begin{figure}[!t]
\includegraphics[width=8.6cm]{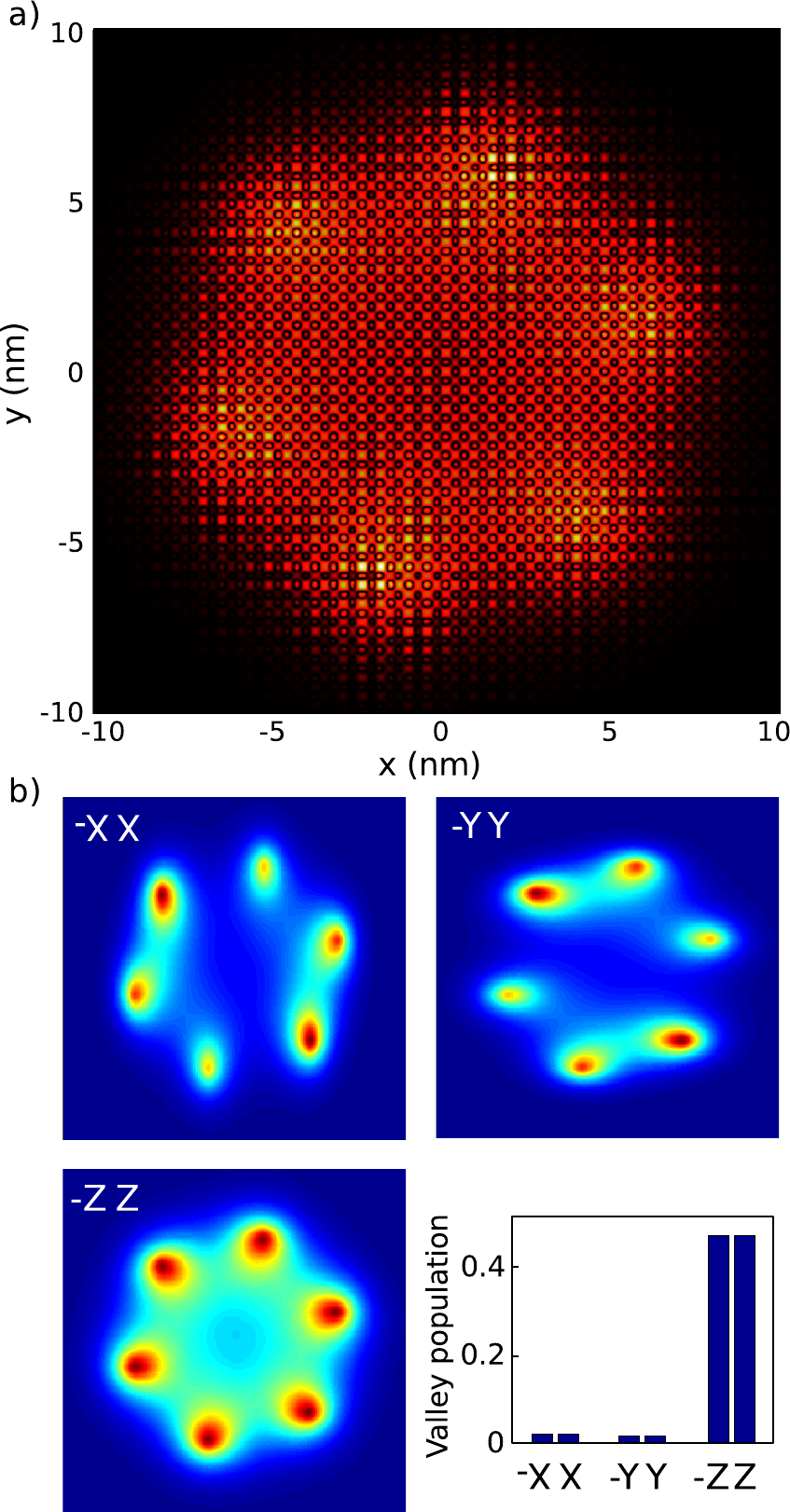}
\caption{a) The ground state electron density for the hexagonal structure formed by six donor atoms with an edge of the hexagon aligned long [110] crystallographic axis and b) corresponding single-valley envelope functions and valley compositions.} 
\label{fig:hex_2}
\end{figure}

As is shown in Figs. \ref{fig:hex_1} and \ref{fig:hex_2}, in both hexagonal structures, the population of Z-valley (perpendicular to the plane) is more pronounced than in the triatomic structure (see Fig. 3). The polarization of Z-valley is even larger in P$_617_{\{110\}}$ . Comparing to the triangle molecule, these examples show that the polarization in the favor of the valleys $\pm$Z  is growing with the number of donors arranged in the plane perpendicular to the main axis of the isoenergetic ellipsoid of the corresponding valley. The limit of large number of coupled donors corresponds to the case of Si:P $\delta$-doped layers. For these structures, a strong Z-valley polarization of lowest states (1$\Gamma$ and 2$\Gamma$ states in the supercell 2D states classification) has been also reported and elucidated by means of the planar Wannier orbitals with empirical pseudopotentials \cite{delta1}, supercell density functional theory \cite{delta2} and supercell tight-binding method \cite{delta}. 

\begin{figure}[t]
\centering
\includegraphics[width=7.3cm]{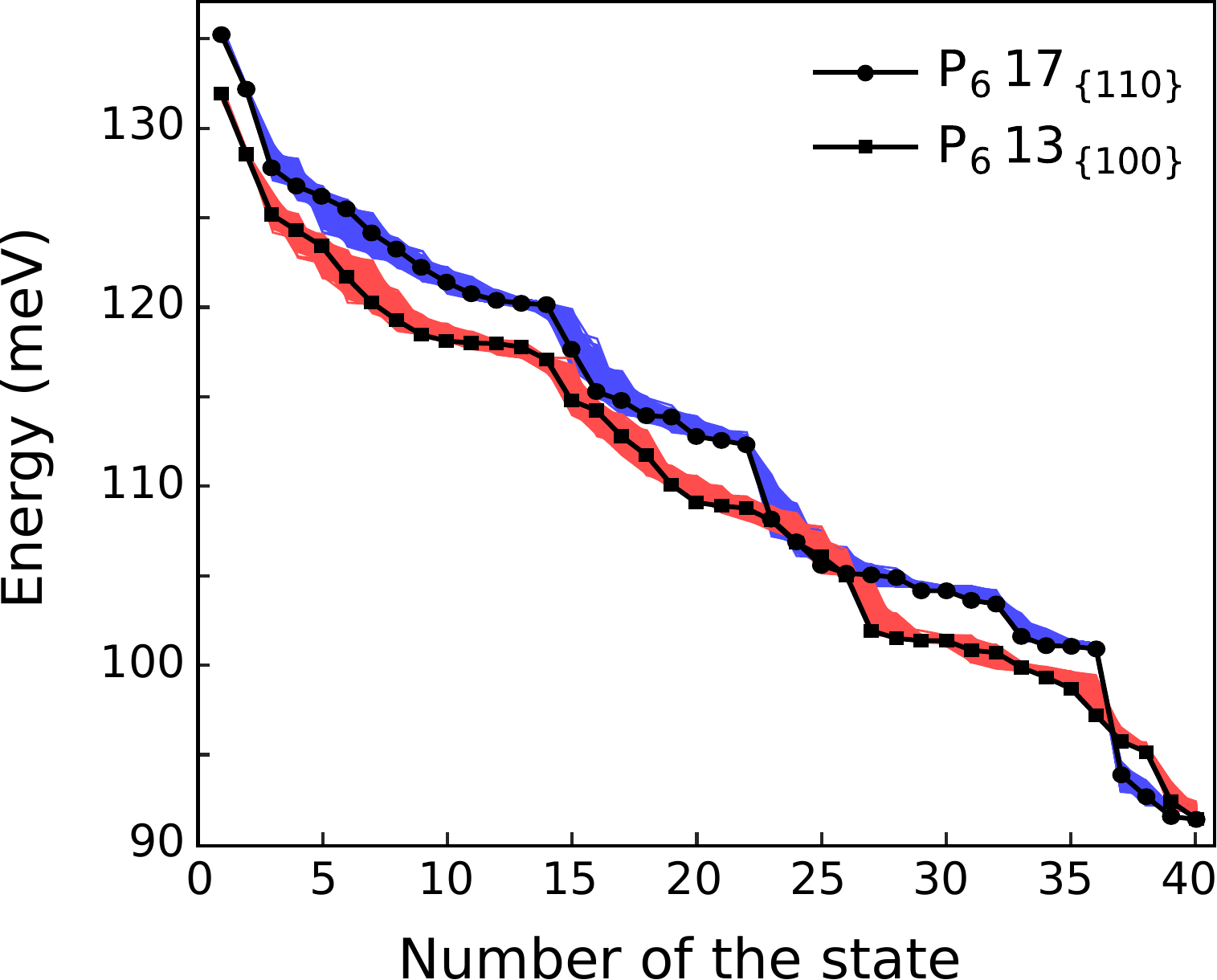}
\caption{The energy spectrum six phosphorus donors arranged in the hexagonal structures} 
\label{fig:P6_spectrum}
\end{figure}

The symmetry of six phosphorus donors arranged in a hexagonal structure have richer valley interference patterns than in the triatomic structures discussed above. This is due to a complex superposition of the six phase factors in Eq. (\ref{nondiag_me_s1}). As a result there is almost no degeneracies in the  energy spectra plotted in  Fig. \ref{fig:P6_spectrum}. The rotation of the hexagonal structure relative to the crystallographic axes changes the energy spectrum within a 5 meV range. The structure P$_617_{\{110\}}$ has a more pronounced band structure (groups of states of a specific symmetry are separated by larger energy gaps). The binding energies are larger in  P$_617_{\{110\}}$ than in P$_613_{\{100\}}$ almost for all energy states. 

For all technological approaches, a deterministic positioning of impurities in silicon crystal lattice is always characterized by a spatial disorder. In the case of the hydrogen-resist lithography, the position of impurities fluctuates within a single lattice constant. To estimate the effect of spatial disorder on single-electron energy spectra of hexagonal impurity molecules we implement the following numerical experiment: 1) first we arrange atoms in a regular structure, 2) then we assign an integer number to all neighboring sites around an impurity, zero is assigned to the impurity site itself, 3) at the next step, the position of each impurity atom remains unchanged or is randomly shifted to one of its neighboring sites in the  $xy$ plane that is defined by the random number generator, 4) having a new arrangement of impurities we compute the energy spectrum, 5) starting from the step one the whole procedure is repeated one thousand times. Each one of the thousand energy spectra is plotted in Fig. \ref{fig:P6_spectrum} as a red thin line for P$_613_{\{100\}}$ and a blue thin line for P$_617_{\{110\}}$. The collections of lines form a band whose width indicates the maximum absolute deviation for each energy state.

The maximum deviation caused by an amount of spatial disorder within a single lattice constant is 3 meV. Such fluctuations of the positions of impurity atoms  do not affect significantly the envelope function, so the energies change mostly because of changes in the valley-orbit coupling which is more sensitive to relative positions of donors. Indeed, the results of the simulation evidence that not all energy levels have same sensitivity to spatial disorder and the sensitivity is determined by the valley composition. For instance, the two lowest states are almost insensitive to small shifts of dopant atoms in the $xy$ plane, because they are formed mainly from Z-valleys.

The electron densities for other energy states of the hexagonal structures P$_617_{\{110\}}$ and P$_613_{\{100\}}$ are reported in Supplementary Information.

\section{Summary}

We have theoretically investigated the single-electron electronic structure  of polyatomic  phosphorus donor molecular structures embedded in silicon. The effects of the effective-mass anisotropy and of the valley-orbit coupling lead to the lifting of the six-fold degeneracy of the conduction band minima observed in the bulk silicon. In a system of several coupled phosphorus donors, the binding energy is growing with the number of donors and with decreasing internuclear distances. In the linear molecules, the energy splitting between the ground state and the first exited state is almost independent of the number of impurities for the range of internuclear distances from 1.5 nm to 6.6 nm. As has been discussed above, at the distances less than 6.6 nm the energy splitting is determined by the effect of the valley-orbit coupling, while at the distances larger than 6.6 nm the splitting is determined by the electronic coupling and long-range electrostatic interaction.

The analysis of the planar molecules made of the three and six coupled donors indicates that the valley composition of the ground state  is polarized in favor of the valleys oriented perpendicular to the plane of the molecule. The polarization is enhanced with increasing the number of atoms forming the planar molecule. Changing the orientation of the six donors arranged in the hexagonal structure relative to the crystallographic axes results in changes of the energy spectra within 5 meV.

In the hexagonal donor arrangements, we have investigated the effect of the spatial disorder that is characterized by an in-plane random displacement of a phosphorus atom within a single unit cell. Such small displacements lead to changes in the electronic energy spectra because in donor molecules, the valley orbit coupling is very sensitive to the relative position of the constituent dopants in the silicon lattice. How much the energy of a given state is affected therefore depends on its valley composition. Our simulation shows that the change in energy of the states that are affected most is less than 3meV, which is smaller than the spacing between the energy levels. The effect of disorder is significantly reduced for the states whose valley population is polarized in favor of the valley whose isoenergetic ellipsoids are aligned along the axis perpendicular to the plane of the molecule.  

\acknowledgments

This work was jointly supported by the proactive collaborative projects TOLOP (318397) and MULTI (317707) of the Seventh Framework Program of the European Commission. The authors acknowledge fruitful discussions with J. Salfi and J. Bocquel. F.R. acknowledges support from Fonds National de la Recherche Scientifique, Belgium, and S.R. acknowledges support from the the ARC DP scheme (DP120101825).

\newpage

\appendix

\renewcommand\thefigure{\thesection.\arabic{figure}} 

\section{Solving single-valley envelope function equations}
\setcounter{figure}{0}  
\begin{figure}[t]
\centering
\includegraphics[width=7.3cm]{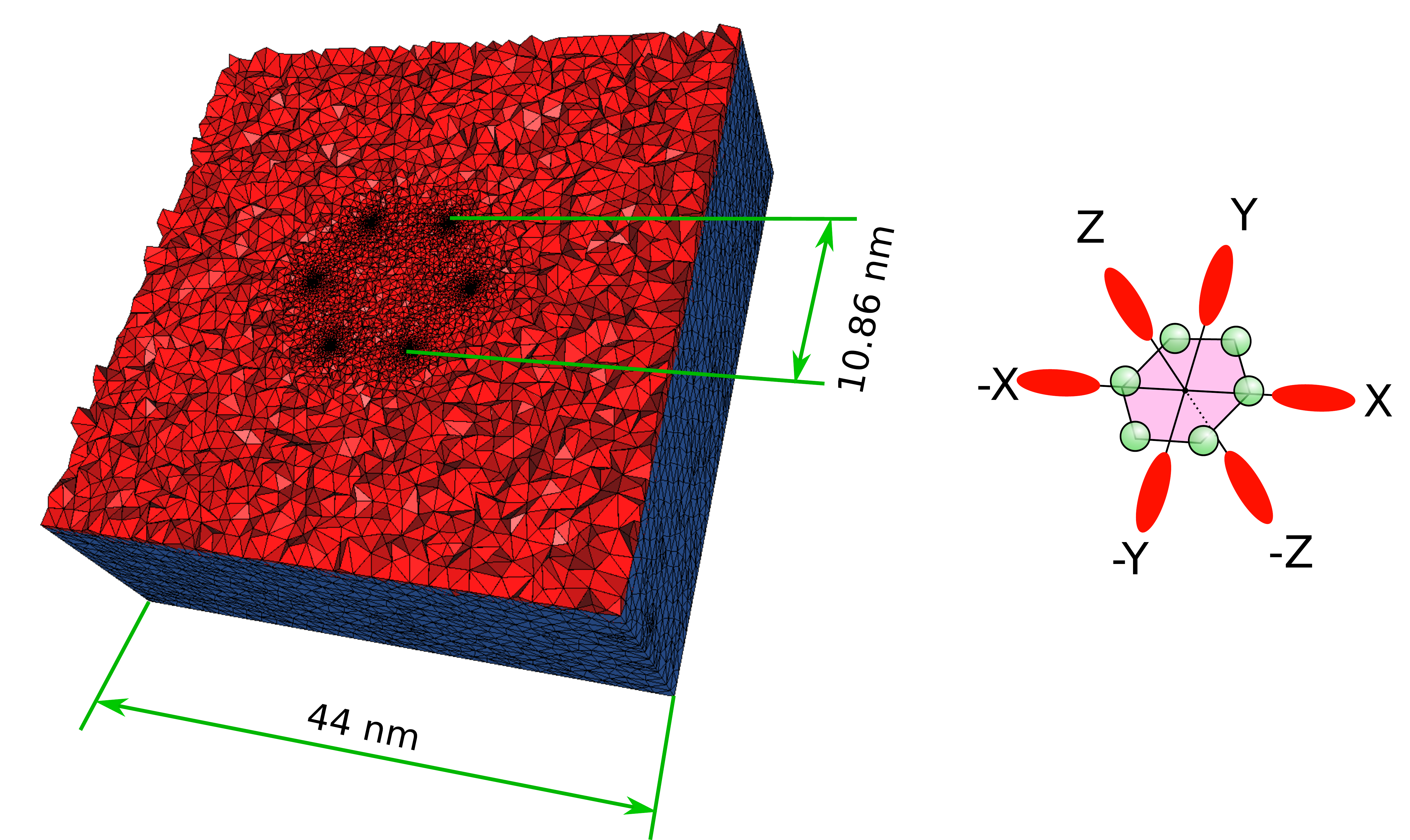}
\caption{a) An example of a generated mesh used in the finite-element-method-based computations for six phosphorus donors arranged in a hexagonal structure, and b) orientations of the bulk silicon isoenergetic ellipsoids relative to the molecular orientation} 
\label{box}
\end{figure}

The single-valley envelope-function equations, representing 3D partial differential equations, have been solved numerically using the finite element method implemented in \textit{FreeFem++} \cite{freefem}. The partial differential equation is subject to Dirichlet boundary conditions: we put the impurities in the middle of a silicon box of cubic shape, and impose $\varphi_{j,\mathbf{k}_0}(\mathbf{r})|_{\Omega}=0$, where $\Omega$ is the boundary surface of the cube. The size of the box has been chosen to ensure an exponential decay of the envelope functions to zero before approaching the box edges.

The mesh has been generated using a 3D tetrahedral mesh generator with a 3D Delaunay triangulator, called \textit{TetGen} \cite{tetgen}, with about 80000 elements. Since the effective potential varies very fast around phosphorus nuclei we have made the adaptation of the mesh density to the function $f\left( \mathbf{r} \right)= \prod_j A \exp(- \alpha|\mathbf{r}-\mathbf{r}_j|^{\beta})$, where $A$, $\alpha$ and $\beta$ are mesh parameters. An example of the grid used in computations is shown in Fig. \ref{box} together with the orientations of the silicon conduction-band isoenergetic ellipsoids relative to the edges of the cube. 

\section{Alignment of substitutional impurity atoms to form a regular hexagon structure}
\setcounter{figure}{0}  
\begin{figure}[t]
\centering
\includegraphics[width=7.3cm]{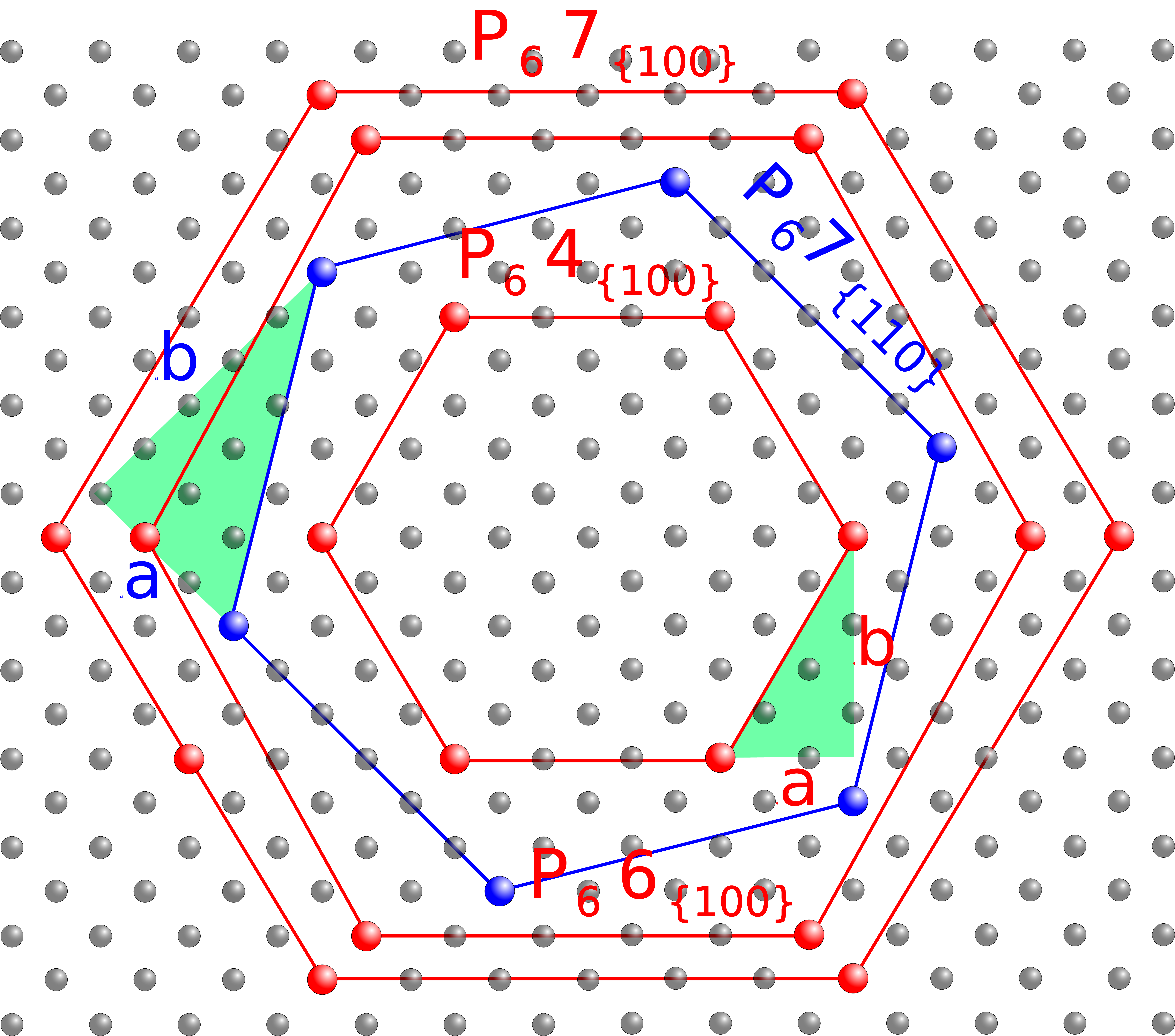}
\caption{Possible alignments of six donor atoms in a regular hexagon structure in the crystallographic plane (001) of silicon (the small grey spheres designate silicon atoms, the large color spheres represent phosphorus impurities). Edge lengths equal $2a$.} 
\label{fig:d6}
\end{figure}

Since silicon has a ​face-centred diamond-cubic crystal lattice an exact arrangement of six substitutional impurities into a regular hexagonal structure is impossible. However, there are many approximate arrangements close to that shape. As a measure of the inaccuracy of each arrangement, we have chosen the ratio between the line segments $b$ and $a$ shown in Fig. \ref{fig:d6}. For a regular hexagon this ratio is equal $\tan \left( 60\degree \right)=\sqrt{3}$. We consider two possible orientations of the hexagonal structure in the plane (001): one has an edge aligned along the axis [100], and other has an edge oriented along the axis [110]. By minimizing $b/a$, we have found a sequence of possible structures collected in Tab. \ref{tab:si}. For each structure, we use a nomenclature described in Sec. III B. The table contains the first twelve possible structures only.

\begin{table}[!h]
\caption{\label{tab:si}$P_6$ hexagonal structures}
\centering
\begin{ruledtabular}
\begin{tabular}{lcccc}
 Structure  & a & b & Edge        & Deviation from \\
            &   &   & length (nm) &  a regular hexagon\\
            &   &   &   &  $\left( ^{b}/_{a}-\sqrt{3} \right) $\\
\hline
    P$_64_{\{100\}}$                    & 1.5 & 2.5 & 1.629      & -0.0654  \\
    P$_66_{\{100\}}$                    & 2.5 & 4.5 & 2.715      & +0.0679  \\
    P$_67_{\{100\}}$/P$_67_{\{110\}}$   & 3   & 5   & 3.258/2.30 & -0.0654  \\
    P$_69_{\{100\}}$/P$_69_{\{110\}}$   & 4   & 7   & 4.344/3.07 & +0.0179  \\
    P$_610_{\{100\}}$                   & 4.5 & 7.5 & 4.887      & -0.0654  \\
    P$_611_{\{100\}}$/P$_611_{\{110\}}$ & 5   & 9   & 5.43/3.84  & +0.0679  \\
    P$_612_{\{100\}}$                   & 5.5 & 9.5 & 5.973      & -0.0048  \\
    P$_613_{\{100\}}$/P$_613_{\{110\}}$ & 6   & 10  & 6.516/4.61 & -0.0654  \\
\end{tabular}
\end{ruledtabular}
\end{table}

In Sec. III B we consider in details two hexagonal structures, P$_613_{\{100\}}$ and P$_617_{\{110\}}$, with approximately equal edge lengths.

\end{document}